\newcommand{\ee} {{\cal E_{\rm q}}}
\newcommand{\p} {{\cal P_{\rm q}}}
\newcommand{\B} {{\cal B_{\rm q}}}
\newcommand{\eL} {{\cal E}}
\newcommand{\pL} {{\cal P}}
\newcommand{\BL} {{\cal B}}
\newcommand{\C} {{\cal C}}
\newcommand{\II} {{\hbox{\tenrm I\kern-.19em{I}}}}

\documentstyle[12pt,a4]{article}
\begin{document}
\pagestyle{empty}

\vspace*{40mm}
{\bf \noindent QUANTUM ALGEBRAS

\vspace{2mm}
\noindent AND LIE GROUPS\footnote [1]
{{\it Contribution to the Symposium
``Symmetries in Science VI" in honor of L.C.Biedenharn

\indent \phantom{}~~Bregenz, Austria, August 2-7, 1992 - B.Gruber ed., Plenum
(New York, 1992)}}}

\vspace{10mm}
\noindent \hspace*{1in} Enrico Celeghini

\vspace{5mm}
\noindent
\hspace*{1in} Dipartimento di Fisica, Universit\`a di Firenze \newline
\hspace*{1in} and I.N.F.N. Sezione di Firenze  \newline
\hspace*{1in} L.go E.Fermi 2, I50125 Firenze, Italy. \newline
\hspace*{1in} e--mail {\footnotesize CELEGHINI@FI.INFN.IT}

\vspace*{2mm}

\begin{flushright}
{\small\em To Larry Biedenharn\\ master and friend}
\end{flushright}
\noindent {\bf INTRODUCTION}\\

Quantum groups are born in Leningrad\cite{Fa} in connection
with the quantum inverse problem method as it arises in soliton theory.
Other motivations came out after and the logic of the
subject has been repeatedly reversed\cite{Dr} and it is still not
univocally established. The approach followed here attempts to use the
link of the field with Lie groups as much as it is possible at the moment
and, perhaps, a little more.

Quantum groups\cite{Re} are groups of matrices with not commuting
elements. Only in simplest cases the quantum algebras quoted in the title
have been shown to be the corresponding differential structures\cite{Wo}.
In general (and often only in principle, because the R-matrices also are
not always known) the connection between quantum groups and quantum algebras
is built using the algebraic approach to the quantum inverse scattering
method i.e. the Yang-Baxter equation.

The first question on quantum algebras is about their general definition.
It is not unreasonable to think that the work of Woronowicz
can be extended
and that, any way, we have a one to one correspondence among  quantum algebras
and quantum group, as it happens in the Lie limit.
If this is true, the two properties universally ascribed to them, i.e. the
fact that they are deformations of Lie
algebras and Hopf algebras, are not enough to define quantum algebras.
{}From the Truini\cite{Tr} contribution to this symposium, it seems indeed
that the class of structures so defined is too wide to hope for
a $q$-group for each of them. In the same time it is not clear
which of the properties we see in concrete examples (primitive Cartan
subalgebra
with unchanged commutators, coalgebra linear in not vanishing roots...) are
really relevant.
 So we are reduced with a temporary
operative definition: quantum algebras
are quantum algebras i. e. the ones we can find in literature.

Let us consider now the r\^ole of Lie groups in physics:
we give a physical meaning to the generators and to some simple
polynomials of them (physical observables) and represent them as operators on
the Hilbert space of physical states. In addition we use a (trivial) rule
to construct the observables of a composed system from the ones of the
components. In more mathematical terms we deal with the algebra, the
representations and the coalgebra.
But all these things are well known for quantum algebras
also:
the ``generators" are not univocally defined, as in the Lie case, because
we have not primitivity to identify them but we can choose them on physical
basis; the representations coincide (at least for simple cases) with the ones
of Lie algebras because the
universal enveloping algebra (UEA) is the same;
the coalgebra acting  on the product space
{}~$V\otimes V$~ and being isomorphic to the algebra on $V$ has the correct
properties to extend to $q$-algebras the product of representations.

Actually Lie algebras have another fundamental property: they can be
integrated to Lie groups by the exponential mapping and nothing similar
exists for quantum algebras at least in commutative structures that can be
realized in a Hilbert
space. But in the following we will attempt to convince you that Lie structures
are so close that we can use them in such a way that we do not really need
anything more to perform with quantum algebras the same play we are used to
do with Lie groups.
Indeed the group has the canonical r\^ole given by the adjoint action i.e.
the group is used to realize infinite duplicates of the algebra, each
corresponding to one ``observer" or to one ``reference frame".
To the algebra, and to the algebra only, we reserve the part to describe
quantities that can be measured. So we (at least I) will be content to
have the same possibility to realize, in similar way, a group of
transformations
on the quantum algebra and to give to the infinite duplicates of the
$q$-algebra so found
the same interpretation as observables in different reference frames.
We shall see that, at least on the representations, the play will be
done exactly by the Lie group, integration of the Lie limit of the
$q$-algebra.

To be more pedagogical let we discuss the argument in general but
referring to an example.
We hope that, in the end, you will admit that, at a formal level
admissible for a physicist :

1) Quantum structures can be dealt more or less as the Lie one.

2) The $q$-coalgebra also plays a fundamental physical r\^ole as it establishes
   the rules for combining single elementary objects into composite systems.
   Its ``generators" have, of course, the properties we require to
   physical observables i.e. they are hermitian and symmetric in the
   two components.

3) The adjoint action of the Lie group can by extended naturally to
   $q$-algebra and $q$-coalgebra.

4) Quantum structures are not, or not entirely,  involved inventions of the
   mind but, exactly like Lie ones,
   express also fundamental invariances of the physical world.
   The ``simple example" we introduce is, indeed, more than a simple example:
   it will
   show that, as in particle physics we define a particle as an unitary
   irreducible representation of the Poincar\'e group,
   in solid state physics we have to define the phonon as an unitary
   irreducible representation of the ``Poincar\'e quantum group" as defined in
   the following.

\vspace*{10mm}
\noindent {\bf LIE GROUP APPLIED TO QUANTUM ALGEBRA }\\

It is well known that the universal enveloping algebra of a quantum algebra
coincide, for $|q|\not= 1$, with the  universal enveloping algebra of the
corresponding Lie
algebra ($UEA^q \equiv UEA$). This means that it is possible\cite{Ch} to
realize a one to one mapping among the quantum ``generators" $X^q_i$ and
the corresponding Lie ones (i.e. the ones obtained in the Lie limit) $X_i$
and vice versa or, in formulas,
$$
X^q_i~ \equiv ~X^q_i(\bar{X})~~~~~~~~~~~~~~ X_i~ \equiv ~X_i(\bar{X^q})
$$
where $\bar{X}$ ($\bar{X^q}$) means the full set of Lie (quantum) generators.
As it is written in textbooks, the action of the group on the Lie
generators is
$$
X'_i~~~ \equiv ~~~{\it g}~ X'_i~ {\it g^{-1}}~~~ \equiv
{}~~~e^{{\it i}\bar{\alpha}\cdot\bar{X} }
{}~X_i
{}~e^{-{\it i}\bar{\alpha}\cdot\bar{X}}
$$
and it saves the Lie commutation relations (CR).
But this mapping generates a ``natural" action of the Lie group
on the quantum generators also.  Indeed, we can define
$$
{X^q_i}'\,\, \equiv \,\,
e^{{\it i}\bar{\alpha}\cdot\bar{X}} X^q_i e^{-{\it i}\bar{\alpha}\cdot\bar{X}}
\,\, =\,\,  X^q_i(
e^{{\it i}\bar{\alpha}\cdot\bar{X}} \bar X e^{-{\it
i}\bar{\alpha}\cdot\bar{X}})
\,\, =\,\,  X^q_i(\bar X')
$$
in analogy with the action on the Lie generators: the quantum CR also
$$
[ X^q_i, X^q_j ]\,\, =\,\, F_{ij}(\bar{X^q} (\bar{X}))
$$
are saved by the mapping:
$$
[ {X^q_i}', {X^q_j}' ]\,\, =\,\, e^{{\it i}\bar{\alpha}\cdot\bar{X}}\,
  [ X^q_i, X^q_j ]\,
e^{-{\it i}\bar{\alpha}\cdot\bar{X}}\,\, =\,\,
{\it F}_{ij}(\bar{X^q} (\bar{X'}))\,\, =\,\, {\it F}_{ij}({\bar{X^q}}')\,.
$$

The relevant point is that this
is still true for the coalgebra also. The coalgebra of the Lie algebra
we are considering is not primitive because it is fixed by the coalgebra
of the quantum algebra and the mapping with the Lie one:

$$
\Delta (X_i)\, \equiv \,X_i(\Delta (\bar{X^q}))\,;
$$
but, because we are dealing with an Hopf algebra, we have still the
homomorphism:
$$
[ \Delta (X_i), \Delta (X_j) ]\,\, =\,\, {\it f}^k_{ij} \Delta (X_k)\,.
$$

So we can repeat the preceding discussion: the same adjoint action of the Lie
group does not change not only the Lie coalgebra CR,
$$
[\Delta (X_i)',\Delta (X_j)']\, =\, {\it f}^k_{ij}\Delta (X_k)'
$$
but also their quantum counterparts,
$$
[ \Delta (X^q_i)', \Delta (X^q_j)' ]\,\, =\,\, {\it F}_{ij}(\Delta
(\bar{X^q})')
$$
where the action of the Lie group has been defined, as usual, by means of the
coproduct
\begin{eqnarray*}
\Delta (X_i)'\,\, &\equiv& \,\,
e^{{\it i}\bar{\alpha}\cdot\Delta (\bar{X})}
\,\Delta (X_i)\, e^{-{\it i}\bar{\alpha}\cdot\Delta (\bar{X})}\,\,
 = \,\,\Delta (X_i')\,\\
\Delta (X^q_i)' &\equiv&
e^{{\it i}\bar{\alpha}\cdot\Delta (\bar{X})}
\Delta (\bar{X^q_i}) e^{-{\it i}\bar{\alpha}\cdot\Delta (\bar{X})}\,
 = \,\Delta (\bar{X^q_i}')\,
\end{eqnarray*}

In such a way, the adjoint action has been fully extended to the $q$-algebra.

To be more clear let us consider our example: $E_q(1,1)$.
 $E_q(1,1)$ is a deformation of $E(1,1)$ and  is generated, as its
Lie counterpart,
by  three generators. In the following we call $\eL, \pL, \BL$
the generators of $E(1,1)$ and $\ee, \p, \B$ the ones of $E_q(1,1)$.
We have:
$$
[\B,\p] = i\ee\,,~\,\,\, [\B,\ee] = (i/w)\, \sinh(w\p)\,,~\,\,\, [\ee,\p] =
0\,.
$$
\noindent where $w \equiv \log q$. The coproducts read
\begin{eqnarray*}
 \Delta (\p)&=& \p\otimes 1\ +1\otimes \p,,\\
 \Delta (\ee)&=& \ee\otimes e^{w\p /2}+e^{-w\p /2}\otimes \ee\,,\\
 \Delta (\B)&=& \B\otimes e^{w\p /2}+e^{-w\p /2}\otimes \B\,;
\end{eqnarray*}

\noindent and the antipodes
$$
\gamma(\B) = -\B+(i/2) ~w \ee\ ,\,\,\quad
\gamma(\p) = -\p\ ,\,\,\quad \gamma(\ee) = -\ee\ .
$$

The Casimir of  $E_q(1,1)$ is
$$
\C\,\, =\,\, \ee^2\, -\, 4/w^2 \sinh^2(w\p/2),
$$
and, because of the Schur lemma, it identifies the irreducible representations
of the UEA\cite{Po}.
It is straightforward to verify  that this quantum algebra, obtained
by contraction from the well known $SU_q(1,1)$ in ref. \cite{Ce}, satisfies
the Hopf algebra axioms and to realize that
the limit\, $w\rightarrow 0$\, gives the Poincar\'e Lie algebra in one
spatial dimension.

Looking to the expression of $\C$ it is now trivial to realize a mapping of
$E_q(1,1)$ on $E(1,1)$ and vice versa
\begin{eqnarray*}
 \eL &=& \ee~~~~~~~~~~~~~~~~~~~~~~~~~~~~~~~~~\ee~ =~ \eL~~~~~~~~~~~~~~~\\
 \pL &=& 2/w ~\sinh(w\p/2)~~~~~~~~~ ~~~\p~ =~ 2/w~ {\rm arcsinh}(w\pL/2)~~~\\
 \BL &=& {\rm sech}(w\p/2) ~~ \B ~~~ ~~~~~~~~~~
\B~ =~ \cosh[{\rm arcsinh}(w\pL/2)] ~~\BL\,,
\end{eqnarray*}
from which the coalgebra associated to the Lie algebra is easy obtained
\begin{eqnarray*}
 \Delta (\pL)&=& 2/w ~\sinh(w\Delta(\p)/2)~ =~
1/w~~(e^{w\p/2}\otimes e^{w\p/2} - e^{-w\p/2}\otimes e^{-w\p/2})\,,\\
 \Delta (\eL)&=& \ee\otimes e^{w\p /2}+e^{-w\p /2}\otimes \ee\,,\\
 \Delta (\BL)&=& {\rm sech}[w\Delta(\p)/2]~ ~ \Delta (\B)\,;
\end{eqnarray*}

where the r.h.s. can be obviously rewritten in terms of $\eL$, $\pL$ and
$\BL$, showing that the coproduct associated by the mapping is not the
primitive one.

We can now operate with the conventional Lorentz boost $e^{i\alpha
\BL}$ to change the reference frame. The formulas for Lie algebra
(and coalgebra) are
written in textbooks; so we report the quantum ones only:
\begin{eqnarray*}
  \ee ' &=& \cosh(\alpha) ~\ee -  2/w ~\sinh(\alpha) ~\sinh(w\p/2)\,,\\
  \p ' &=& 2/w ~~{\rm arcsinh}\{w/2[ 2/w ~\cosh(\alpha) ~\sinh(w\p/2) ~-
          ~ \sinh(\alpha) ~\ee]\}\,,
\end{eqnarray*}
and, of course,
$$
\Delta(\ee) ' =  \Delta(\ee')\,,~~~~~~~~~~~~~~~~~~
  \Delta(\p) ' = \Delta(\p')\,.
$$

Let look now at the representations. Because we are interested in the
representation with ~$\C = 0$~, let we consider it only.
{}From the irreducible representations
for $E(1,1)$ in the plane-wave basis  with $\C$
and $\pL$ diagonal, the action of quantum variables is obtained as:
\begin{eqnarray*}
\C \,|0,p> &=& 0\,,\\
\p \,|0,p> &=& p ~~|0,p>\,,\\
\ee \,|0,p> &=& E ~~|0,p>,~~~~~\left( E = 2/w ~{\rm sinh}(w p/2)\right) .
\end{eqnarray*}
The full representation
is generated by the Lorentz boost
$e^{-i\alpha \BL}$:
$$
|0, p'>~ =~ e^{-i\alpha \BL} ~|0,p>
$$
and we have
\begin{eqnarray*}
\p |0,p'> &=& p' ~~|0,p'>~~~~~~~~(~p' ~=
{}~2/w ~{\rm arcsinh}[e^{-\alpha} \sinh(w p/2)]~)\\
\ee |0,p'> &=& E' ~~|0,p'>~~~~~~~~( E' = e^{-\alpha} E )
\end{eqnarray*}
and the Lie normalization
become
$$
<0,p'|0,p>~~ =~~ 4/w ~\tanh(w p/2) ~\delta (p'-p).
$$

Discrete transformations act in the same way of the Lie case and will not
be discussed.

\vspace*{10mm}
\noindent {\bf PHONONS}\\

To have directions about  physical applications of $q$-algebras, we must look
to
physical applications of Lie groups, in outline they are applied:

in phenomenology and a good example can be $SU(3)$
(but, from it, we arrived to standard model, quite more that phenomenology...);

in dynamics (think, for instance, to $SO(4)$ and the hydrogen atom...) and

in kinematics as the Poincar\'e group.

\noindent $Q$-algebras applications also can, indeed, be classified in the
same three fields; we have:

papers in phenomenology (for instance the Biedenharn's studies on
$a_q$ and $a^{+}_q$) where $q$ is a new parameter to be fitted\cite{Bi} ;

papers in dynamics (the best example is, to my knowledge, the XXZ model
obtained as deformation of the XXX one, where $q$ measures the breaking
of the rotational symmetry\cite{Pa}) and

applications to kinematics, object of this talk, in which $q$
is connected to lattice spacing\cite{Bo}.

Always we had to find the good ``generators" and give them a physical
meaning.

Roughly speaking $q$-algebras differ from Lie ones
for some exponential of the generators of the kind
$e^{w X}$ (where $w \equiv \log q$). In the usual representations, where
$X$ is a differential operator,
$e^{w X}$ is a finite difference operator: so $q$-groups seem related to
finite difference problems and this suggests to look for applications in
discrete physics.
But, now, we have a problem: $e^{w X}$ must be a number and so $w X$;
in consequence $[w] = [X]^{-1}$, but for simple groups CR are inhomogeneous,
and, so, generators and consequently $w$ (or $q$) must be dimensionless
and standard discrete physics have not dimensionless parameters. On the
contrary
solid state physics consider parameters of the dimension
of a length: the way out are inhomogeneous $q$-algebras; their CR are,
indeed, homogeneous in the abelian subalgebra that can acquire (and actually
acquires) a dimension of $[l]^{-1}$ so that $[w] = [l]$.

{}From a technical point of view, inhomogeneous $q$-algebras have a strange
peculiarity: contrary to their Lie limits, they cannot be obtained by
standard contraction from simple algebras, where by standard we
mean contraction with respect to a subalgebra.
So to build inhomogeneous $q$-algebras from homogeneous ones
we are forced to define a new kind of contraction that does not save
a subalgebra and involves, in the limit, the quantum parameter $w$ also,
using, in such a way, a same sort of analyticity in $w$.
The details can be found in\cite{Ce} , where, among other things,
the contraction of
$SU_q(1,1)$ to  $E_q(1,1)$ is explicitly performed.

As we shall show, the kinematical invariance of crystal
is, indeed, described by this Poincar\'e quantum group.
In front of space-time of special relativity (and its mathematical
description, the Poincar\'e Lie group), crystals (and Poincar\'e
quantum group) have not in themselves a continuous translational
invariance but only a discrete translational invariance as discussed
before.

Let us so consider the linear chain of equal masses lying at a distance $a$
{}from one another, with nearest neighbor harmonic interaction.
The equations of motion are:
$$
\ddot{z}_j(t)=\omega^2 \left(z_{j-1}(t)+z_{j+1}(t)-2z_j(t)\right)
$$
where $z_j(t)$ is the displacement of the $j$-th mass $(j=0,1,\dots, N)$.
Periodic boundary conditions are assumed
and initial conditions $z_j(0),\ \dot{z}_j(0)$ must be specified.

We embed the ordinary system for displacements into the partial differential
equation (PDE)
$$ \left(\partial_t^2 + (2v/a)^2 \sin^2(-ia\partial_x/2)
\right)\, z(x,t)=0\ ,
$$
where $v=\omega a$.
The periodic conditions are $z(0,t)=z(Na,t)$ while
the Cauchy data consist in the assignment of smooth functions $z(x,0)$
and $\partial_t z(x,0)$. When $z(j a,0)=z_j(0),\ \partial_t z(j a,0)=
\dot{z}_j(0)$ for all $j\,$,  it is easy to see that the solutions
of the ordinary system are directly obtained as $z_j(t)=z(j a,t)$
irrespectively of the
behaviour of the solutions in the points $x\not = ja$.

The continuum limit $a\rightarrow 0$\, of PDE obviously reproduces
the Klein--Gordon equation
in dimension $(1+1)$ with velocity $v$ and mass $m = 0$.
This constitutes a differential realization of the
Casimir of the $E(1,1)$ algebra, which is actually the kinematical symmetry
of the continuous system. Likewise, our PDE identifies
a realization with Casimir $\C = 0$ of the pseudoeuclidean quantum algebra
$E_q(1,1)$ described before which in its own right can be
considered the kinematical symmetry of the harmonic crystals.
The dimensional deformation parameter $w$, also, has a simple physical
interpretation as it is related to lattice spacing by $w =i a$.

Because $w$ results imaginary, $q$ is on the unit circle:
the topology of $E_q(1,1)$ is now completely
different but the essential results about the connection with Lie groups are
still valid.

The related realization of the $q$-algebra is obtained from ref.\cite{Ce}
yielding
$$\ee=(i/v)\ \partial_t\,, \quad ~~  \p= - i \partial_x\,, \quad ~~
\B=i(x/v)\partial_t-(vt/a)\sin(-ia\partial_x)\,.$$
In the momentum representation, a realization of the $E_q(1,1)$
in terms of the diagonal $\p$ and the position operator
$X=i \partial/\partial p$ is given by:

\begin{eqnarray*}
 \ee &=& (2/a)\ \sin(ap/2),\\
\B  &=& (1/a)\ \bigl\{\sin(ap/2),X\bigr\}_+\ ,\\
\p   &=& p\ .
\end{eqnarray*}

Our notation has been chosen for its transparent
physical meaning, but its is not the good one from mathematical point
of view. As a matter of fact, mathematicians do not introduce $\p$ but
$k = e^{i a\p}$. If we rewrite everything in terms of $k$ the real topology
appears and we see that $\p$ is determined up to an integer multiple of
$2\pi /a$. In such a way, the topology of the $q$-algebra with $|q|=1$
implies what in solid state physics is called the reduction to
the first Brillouin zone i.e. the limitation of the values of $\p$ and $\ee$:
$ 0\leq p<2\pi/a, \ee>0 $.

The expression for the generator $\B$ can be inverted in $X$:
$$ X=(1/2)\ \bigl\{\ee^{-1},\B\bigr\}_+\ .$$
The time derivative of $X$ is given by $\dot{X}=iv\ [\ee,X]$ and the
commutator, evaluated from the $q$-algebra, gives the well known
group velocity of the phonons
$$\dot{X} = v_g= v\ \cos(a\p/2)\ .$$

Let us show how the coproduct can be brought to bear to study
composed systems and, in particular, the fusion
of phonons. It is well known that, when the symmetry is given by a Lie algebra,
the generators of the global symmetry of a composed system are obtained by
summing the generators of the symmetry of the elementary constituents. This
is related to the fact that
each generator $G$ of a Lie algebra is a primitive element,
{\it i.e.} $\Delta(G)=\II \otimes G + G\otimes \II \ $.
Then $G^{(1)}\equiv G\otimes \II \ $ acts
on the vector space of the first elementary system
and $G^{(2)}\equiv\II \otimes G$ on the second. The algebras generated by
$G^{(1)}$ and $G^{(2)}$ are both isomorphic to that generated by $G$ and since
$\Delta$ is a homomorphism of algebras, then $G^{(1)}+ G^{(2)}$
generates the same symmetry on the composed system.
In the quantum group context we can have non primitive generators, but
the very same considerations are still valid, after that the symmetries both
of pseudo-particles and of $q$-algebras have been considered.

Phonons are bosons: to save their statistics and generate a
correct composite system,  we need symmetrical operators in $V\otimes V$.
So the operators of the coalgebra  we have found cannot describe the
observables of the two phonons system: they are not symmetric (and not even
hermitian). In the same time $q$-algebras
have always a symmetry in themselves:  $q \leftrightarrow q^{-1}$ (in our
case equivalent to $a \leftrightarrow -a$) corresponding
to the exchange of the two spaces in $V\otimes V$. So we can
attempt to impose this symmetry by
hand, substituting all the
coalgebra with its symmetrized form under the transformation
$a \leftrightarrow -a$.
\begin{eqnarray*}
\p^s &=& \p^{(1)} + \p^{(2)}\\
\ee^s &=& \cos\left(a\p^{(1)}/2\right)\,
\ee^{(2)}+\cos\left(a\p^{(2)}/2\right)\,
\ee^{(1)}\\
\B^s &=& \cos\left(a\p^{(1)}/2\right)\, \B^{(2)}+\cos\left(a\p^{(2)}/2\right)\,
\B^{(1)}\,.
\end{eqnarray*}
Now $\p^s$, $\ee^s$ and $\B^s$ are all
symmetric and hermitian and an explicit check shows that they still
close the $E_q(1,1)$ algebra.
They are, in such a way, good candidates to describe the observables for the
system of two phonons.
It must be stressed, in the same time, that they are not
a coalgebra of our $E_q(1,1)$ because they do not satisfy to all the
requirement
of an Hopf algebra; in particular they do not allow for building, by
iteration, the operators for three and more phonons:
the global operators must be calculated from the original coproduct up to
the required number of phonons and then completely symmetrized. It is easy to
show that for each $n$ this procedure close again the $E_q(1,1)$ algebra.

Let us look, now, how the simplest composed system i.e. a phonon obtained
by fusion of two phonons is described by the $q$-algebra.
Because $\p$ is defined up to $2\pi /a$,
the composition of the momenta also has the same property:
 $\p^s = \p^{(1)} + \p^{(2)} + 2\pi n/a$,
showing that the Umklapp process is implied by the quantum group symmetry.
The other global generators, that depend from the $\p$'s through
trigonometrical functions only, have not this problem.

In concrete, take
two differently polarized phonons with the same direction of propagation,
velocity parameters $v_1$ and $v_2$ and dispersion relations\cite{As}:
$$\Omega_1 = (2 v_1/a) \sin(a\p^{(1)}/2), \quad ~~~
\Omega_2 =(2 v_2/a) \sin(a\p^{(2)}/2)\ ,$$
where $\Omega_1 = v_1 ~\ee^{(1)}$ and  $\Omega_2 = v_2 ~\ee^{(2)}$ are
the energies of the two phonons.
The explicit coproduct of $\ee$ reads
$$
\ee^s=\cos\left(a\p^{(1)}/2\right)\, \ee^{(2)}+\cos\left(a\p^{(2)}/2\right)\,
\ee^{(1)}=
\ (2/a)\sin\left(a(\p^s)/2\right) \ .
$$
To realize the fusion,
the energy conservation implies the existence of a branch of the
dispersion relation with velocity $v$ such that the global energy
$\Omega=\Omega_1+\Omega_2$ is related to
$\ee^s$ by $\Omega= v|\ee^s|$ .

Moreover, from the same definition of the position operator in terms of the
algebra, we obtain the position operation of the two phonon system:
$$ X^s=\frac{1}{2} (X^{(1)}+X^{(2)}) + \frac{1}{2}
\biggl\{\frac{\sin(a(\p^{(1)}-\p^{(2)})/2)}{\sin(a(\p^{(1)}+\p^{(2)})/2)},
\frac{1}{2} (X^{(1)}-X^{(2)})\biggr\}_+\ ,
$$
which reproduces the Heisenberg algebra $\ [X^s,\p^s]=i\ $ for the global
variables. Finally the group velocity of the composite system
$\dot{X^s}=i\ [\Omega,X^s]=v\ \cos(a\p^s/2)\ $
appears formally identical to that of the elementary system, having performed
the Umklapp process.

To conclude let we stress that there is nothing of peculiar in the phonon,
the relevant point being its dispersion relation connecting energy and
momentum: all the quasi-particles with the same dispersion relation
can be handled with $E_q(1,1)$. Analogously the
quasi-particles with dispersion relation of the kind:
$\ee \approx \sin^2(a \p/2)$
are unitary irreducible representations of the Galilei quantum group
$\Gamma_q(1)$ as can be seen in the contribution of Tarlini\cite{Ta}.

\end{document}